\documentclass[twocolumn]{aastex631}

\usepackage{float}
\usepackage{amsmath}
\usepackage{afterpage}

\submitjournal{ApJ}

\shorttitle{Isolated Stellar-Mass Black Hole}
\shortauthors{Sahu et al.}

\begin{document}

\newcommand{\Gaia}{{\it Gaia}}
\newcommand{\Hipp}{{\it Hipparcos}}        
\newcommand{\Hubble}{{\it Hubble Space Telescope}}
\newcommand{\HST}{{\it HST}}
\newcommand{\IUE}{{\it IUE}}
\newcommand{\JWST}{{\it JWST}}
\newcommand{\kms}{{\>\rm km\>s^{-1}}}
\newcommand{\Mlens}{M_{\rm lens}}
\newcommand{\Spitzer}{{\it Spitzer}}
\newcommand{\Lya}{Ly$\alpha$}
\newcommand{\Teff}{T_{\rm eff}}
\newcommand{\OB}{OGLE-2011-BLG-0462}
\newcommand{\phil}{\hbox{$\varphi_{\mathrm{LS}}$}}
\newcommand{\msun}{$M_\odot$}
\newcommand{\amplification}{{magnification}}
\newcommand{\amplified}{magnified}

\newcommand{\DL}{D_{\rm L}}
\newcommand{\DS}{D_{\rm S}}
\newcommand{\masyr}{\rm mas\, yr^{-1}}
\newcommand{\piE}{$\pi_{\rm E}$}
\newcommand{\piLS}{\pi_{\rm LS}}
\newcommand{\tE}{t_{\rm E}}
\newcommand{\thetaE}{\theta_{\rm E}}
\renewcommand{\vec}[1]{\pmb{#1}}

\title{OGLE-2011-BLG-0462: An Isolated Stellar-Mass Black Hole Confirmed Using New  \emph{HST} Astrometry and Updated Photometry }

\correspondingauthor{Kailash C. Sahu}
\email{ksahu@stsci.edu}


\author[0000-0001-6008-1955]{Kailash C. Sahu}
\affil{Space Telescope Science Institute, 3700 San Martin Drive, Baltimore, MD 21218, USA}
\affiliation{School of Natural Sciences, Institute for Advanced Study, 1 Einstein Drive, Princeton, NJ 08540, USA}
\affil{European Southern Observatory,
Karl-Schwarzshild-Stra\ss e 2,
85748 Garching bei M\"unchen,
Germany}
\affil{Eureka Scientific Inc., 2452 Delmer Street, Suite 100, Oakland, CA 94602, USA}

\author[0000-0003-2861-3995]{Jay Anderson}
\affil{Space Telescope Science Institute, 3700 San Martin Drive, Baltimore, MD 21218, USA}

\author{Stefano Casertano}
\affil{Space Telescope Science Institute, 3700 San Martin Drive, Baltimore, MD 21218, USA}

\author[0000-0003-1377-7145]{Howard E. Bond}
\affil{Department of Astronomy and Astrophysics, 525 Davey Laboratory, Penn
State University, University Park, PA 16802, USA}
\affil{Space Telescope Science Institute, 3700 San Martin Drive, Baltimore, MD 21218, USA}

\author[0000-0002-3202-0343]{Martin Dominik}
\affil{University of St Andrews, Centre for Exoplanet Science, SUPA School of Physics \& Astronomy, North Haugh, St Andrews, KY16 9SS, UK}

\author[0000-0002-0882-7702]{Annalisa Calamida}
\affil{Space Telescope Science Institute, 3700 San Martin Drive, Baltimore, MD 21218, USA}

\author[0000-0003-3858-637X]{Andrea Bellini}
\affil{Space Telescope Science Institute, 3700 San Martin Drive, Baltimore, MD 21218, USA}

\author[0000-0002-1793-9968]{Thomas M. Brown}
\affil{Space Telescope Science Institute, 3700 San Martin Drive, Baltimore, MD 21218, USA}

\author[0000-0001-7113-2738]{Henry C. Ferguson}
\affil{Space Telescope Science Institute, 3700 San Martin Drive, Baltimore, MD 21218, USA}

\author[0000-0002-6577-2787]{Marina Rejkuba}
\affil{European Southern Observatory,
Karl-Schwarzshild-Stra\ss e 2,
85748 Garching bei M\"unchen,
Germany}


\begin{abstract}

The long-duration Galactic-bulge microlensing event OGLE-2011-BLG-0462 produced relativistic astrometric deflections of the source star, which we measured using Hubble Space Telescope (\HST) observations taken at eight epochs over $\sim$6~years. Analysis of the microlensing lightcurve and  astrometry led our group (followed by other independent groups) to conclude that the lens is an isolated stellar-mass black hole (BH)---the first and only one unambiguously discovered to date. There have now been three additional epochs of \HST\/ observations, increasing the astrometric time baseline to 11~years. Additionally, the ground-based OGLE data have been updated. We have reanalyzed the data, including the new \HST\/ astrometry, and photometry obtained with 16 different telescopes. The source  lies only $0\farcs4$ from a bright neighbor, making it crucial to perform precise subtraction of its point-spread function (PSF) in the astrometric measurements of the source. Moreover, we show that it is essential to perform a separate PSF subtraction for each individual \HST\/ frame as part of the reductions. Our final solution yields a lens mass of $7.15\pm0.83\, M_\odot$. Combined with the lack of detected light from the lens at late \HST\/ epochs, the BH nature of the lens is conclusively verified. The BH lies at a distance of $1.52\pm0.15$~kpc, and is moving with a space velocity of $51.1\pm7.5 \kms$ relative to the stars in the neighborhood. We compare our results with those of other studies and discuss reasons for the differences. We searched for binary companions of the BH at a range of separations but found no evidence for any.

\end{abstract}

\received{Jan 18, 2025}
\revised{March 6, 2025}
\accepted{March 7, 2025}

\section{Introduction} 

\subsection{First Unambiguous Detection of an Isolated Stellar-Mass Black Hole \label{subsec:firstblackhole} }

Until recently, all of the more than two dozen stellar-mass black holes (BHs) discovered in our Galaxy, and the over 150 detected in external galaxies through mergers that emitted gravitational radiation, were in binary systems. 
The first unambiguous detection of an {\it isolated\/} stellar-mass BH was reported by \citet[][hereafter S22]{Sahu2022}. They showed that the long-duration ($t_{\rm E} \simeq 270$~days), high-amplification ($A_{\rm max} \simeq 400$) microlensing event MOA-2011-BLG-191\slash OGLE-2011-BLG-0462 
(hereafter \OB)
was caused by a dark lens with a mass so large that it must be a BH\null.   {\it Hubble Space Telescope\/} (\HST\/) images obtained at 8 epochs extending over 6 years clearly revealed a relativistic astrometric deflection of a background star's apparent position as the lens passed in front of it.  Photometry of \OB\ 
carried out with 16 different telescopes displayed the parallactic signature of the Earth's motion in the microlensing lightcurve. Spectroscopic observations taken  with large telescopes at the time of peak amplification were used to determine atmospheric parameters of the source star and thus infer its distance.  

The astrometric deflections from the \HST\ data, the parallax from the photometry, and the source distance measurements by S22 were used to determine a lens mass of $7.1\pm1.3\,M_\odot$ and a distance of $1.58\pm0.18$~kpc.   
The lens was shown to emit no detectable light. This, along with the mass measurement, led to the conclusion that the lens is a BH\null. S22 further noted that the proper motion (PM) of the lens is offset from the mean of Galactic-disk stars at similar distances by an amount corresponding to a relative transverse velocity of $\sim$$45\,\kms $. This suggested that the BH had received a ``natal kick'' during its birth in a supernova explosion.

\subsection{Subsequent Developments}

In an independent analysis of the data,
\citet[][hereafter L22]{Lam2022} found a lower mass for \OB\ than the mass derived by our group (S22), reporting a possible range of masses between 1.6 and $4.4\,M_\odot$ and implying 
that the lens could be either a neutron star or a BH\null. 
However, \citet[][hereafter M22]{Mroz2022} re-reduced and improved the ground-based OGLE measurements, and then reanalyzed the astrometric and photometric datasets. M22 found that the astrometric fit residuals in the L22 analysis are larger in the R.A. direction than
in decl., leading to the conclusion that the
L22 astrometric reductions had been affected by systematic errors. 
M22 combined their updated photometry with the S22 astrometry to obtain a mass determination of $7.88\pm0.82\,M_\odot$, statistically consistent with the S22 results, and reaffirming the BH nature of the lens.

Subsequent to these two studies, three additional epochs of \HST\/ observations of \OB\ have been obtained, increasing the number of epochs from 8 to 11. 
\citet[][hereafter L23]{Lam2023} presented a new analysis of the data, including the new \HST\ measurements. They reported a lens mass of $6.03^{+1.19}_{-1.04}\,M_\odot$, 
superseding their earlier study (L22) and finding an agreement within 1-$\sigma$ with the S22 and M22 mass estimates.

\subsection{Our New Analysis \label{subsec:updated} }

Astrometry of the \HST\/ frames of \OB\ is complicated by the presence of a bright neighbor of the source star, lying only $0\farcs4$ away. This neighbor is nearly 20 times brighter than the source at baseline. Thus, images of the source are overlain by the wings of the point-spread function (PSF) of the bright neighbor. These must be subtracted extremely precisely in order to carry out the astrometry, making it crucially important to understand the PSF structure. Moreover, the PSF varies even during a single \HST\/ visit, due to the varying thermal environment as the observatory orbits the Earth (telescope ``breathing''). These effects were fully taken into account by S22 in their analysis of the first eight epochs of \HST\/ observations. 

In this paper, we will update the analysis of S22 by applying the same astrometric analysis techniques to the three new epochs of \HST\/ observations, extending the time baseline from 6 to 11~years and improving the PM measurements for all stars in the reference frame, as well as for the source itself. We will use the rereduced OGLE photometry 
from M22 
in our new analysis. Additionally we will re-examine the distance to the source by considering more carefully its foreground interstellar reddening. 
Our revised analysis, with the additional \HST\/ observations and updated photometry included, leads to results which have higher accuracy, but are consistent with our previous measurements and our conclusion that the lens is a stellar-mass BH\null. We will also examine the data to place limits on any stellar companions of the BH.

\section{New and Updated Data}

In this updated analysis, we include new \HST\/ astrometric observations, as well as the revised  OGLE dataset, which has the longest time coverage and highest photometric accuracy among the available ground-based data.

\subsection{New {\rm HST} Observations}

All of the \HST\/ observations of the \OB\ field have been obtained with the UVIS (ultraviolet and visible light) channel of the Wide Field Camera~3 (WFC3).
The S22 analysis was based on WFC3 images obtained at eight epochs, starting shortly after the closest approach of the lens to the source in 2011, and ending in 2017. Since then, there have been three further WFC3 image sets obtained, one in 2021 and two in 2022. Table~\ref{table:journal} lists the details of all the 11 \HST\/ visits. Two filters have been used: F606W (``$V$'' band) and F814W (``$I$'' band). In the discussion below, the observation epochs are denoted E1 through E11. 
The telescope pointings at each epoch were dithered between individual exposures, to improve sampling of pixel phase space and also to guard against any small-scale local distortions in the detector. 

\begin{deluxetable*}{clcccccc}
\tablecaption{Log of \HST\/ Wide Field Camera 3 Observations of \OB\ \label{table:journal} }
\tablehead{
\colhead{Epoch}           &
\colhead{Date} &
\colhead{HJD$-$2450000}  & 
\colhead{Year}  &
\colhead{Proposal} &
\colhead{Orient} &
\colhead{No.\ Frames}  & 
\colhead{No.\ Frames}   \\
\colhead{} &
\colhead{}         & 
\colhead{}      & 
\colhead{}         &
\colhead{ID\tablenotemark{a}} & 
\colhead{(Deg)}&
\colhead{in F606W\tablenotemark{b}}    &
\colhead{in F814W\tablenotemark{b}}          
}
\startdata
E1 & 2011 Aug 8  & 5782.2 & 2011.600 & GO-12322 & 270.0 & 4 & 5  \\
E2 & 2011 Oct 31 & 5865.7 & 2011.829 & GO-12670 & 276.1& 3 & 4   \\
E3 & 2012 Sep 9  & 6179.7 & 2012.689 & GO-12670 & 269.5& 3 & 4   \\
E4 & 2012 Sep 25 & 6195.8 & 2012.733 & GO-12986 & 271.3& 3 & 4   \\
E5 & 2013 May 13 & 6426.3 & 2013.364 & GO-12986 & 99.9 & 3 & 4   \\
E6 & 2013 Oct 22 & 6587.7 & 2013.806 & GO-13458 & 274.6& 3 & 4   \\
E7 & 2014 Oct 26 & 6956.6 & 2014.816 & GO-13458 &275.2&  3 & 4  \\
E8 & 2017 Aug 29 & 7995.2 & 2017.660 & GO-14783 & 268.3& 3 & 4   \\
E9 & 2021 Oct 1 & 9488.8 & 2021.749 & GO-16760 & 272.0& 5 & 6 \\
E10 & 2022 May 29 & 9729.3 & 2022.407 & GO-16716 & 107.9& 0 & 2 \\
E11 & 2022 Sep 13 & 9835.8 & 2022.699 & GO-16760 & 269.9& 5 & 6 \\
\enddata
\tablenotetext{a}{The PI for GO-16760 was C.~Lam. K.C.Sahu was the PI for all other programs.}
\tablenotetext{b}{Individual exposure times ranged from a minimum of 60~s at E1, to a maximum of 407~s at later epochs.}
\end{deluxetable*}

\vskip 1cm
\subsection{Revised OGLE Photometry \label{subsec:revised_ogle} }

S22 showed that there is a correlation between  \phil, the position angle of the direction of motion of the lens with respect to the source, and \piE, the microlensing parallax parameter, and that the photometric errors were underestimated.  As a result, the value of \phil\ determined from photometric data alone was ambiguous. Addition of astrometry helped in an unambiguous determination of \phil\ and \piE.  This was further investigated by M22, who found that the photometric errors were indeed underestimated, and that the photometric measurements collected under different seeing conditions give systematically different values of \phil. To minimize the errors and systematic effects, M22 rereduced the OGLE difference-imaging photometry using a reference image constructed from a frame taken in the best seeing conditions (with a FWHM of 2.95 pixels, or $0\farcs77$). They also noted that data taken during the commissioning period of the OGLE-IV camera, in the first half of 2010, had large residuals, and these were discarded. 

\section{Fine Points of High-Precision \HST\/ Astrometry \label{sec:astro_measures} }

\subsection{Precise Subtraction of a Time-Varying and Highly Structured PSF in a Crowded Field}

\begin{figure*}
\begin{center}
\plotone{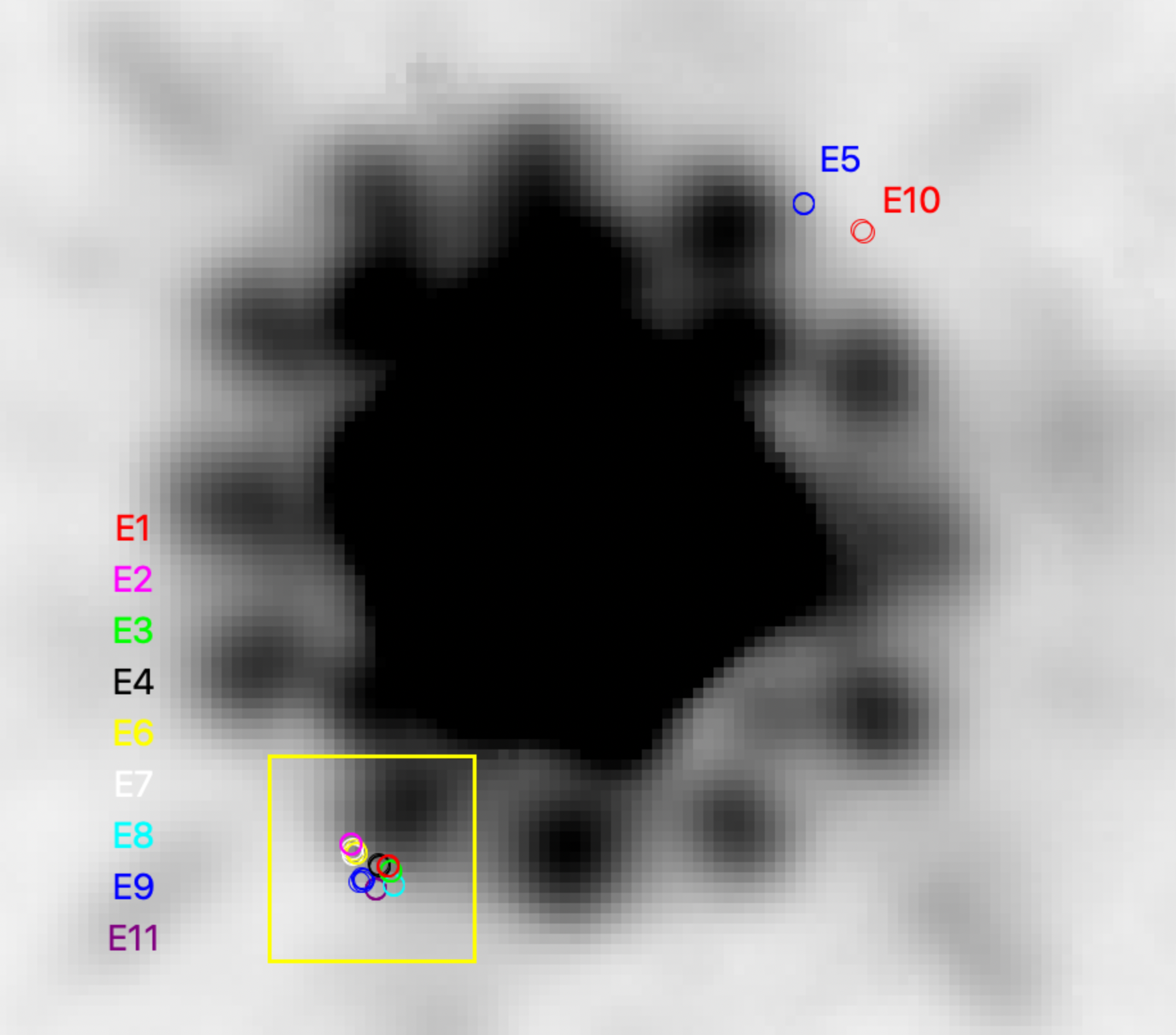}
\end{center}
\caption{
This frame, $1\farcs1 \times 1\farcs0$, shows the PSF in detctor space for isolated stars in WFC3 F814W frames, averaged over all 11 epochs of data. Pixels have been oversampled by a factor of 4, and the stretch has been optimized to show the structure of the PSF in its outer wings. Small open circles mark the locations of the \OB\ source relative to its bright neighbor star at each observation epoch, color-coded as indicated in the figure. The source at epochs E5 and E10 is on the opposite side from the others due to a difference of about $180^\circ$ in telescope roll angle. The yellow square corresponds to the $5\times5$ native WFC3 pixel box used for positional fitting of the PSF.
 \label{fig:dynamic_psf}
}
\end{figure*}

As noted in Section~\ref{subsec:updated}, precise astrometry of the \OB\ source in the \HST\/ frames is challenging because its image is superposed by the PSF wings of a neighboring bright star lying only $0\farcs4$ (10 native WFC3 pixels) away. This is illustrated in
Figure~\ref{fig:dynamic_psf}, showing the highly structured wings of the PSF in the vicinity of the source stars's location. It is thus crucial that PSF subtraction is done correctly in the astrometric reductions. 

In  Figure~\ref{fig:dynamic_psf} the original WFC3 pixels have been oversampled by a factor of 4.
Centered in the frame is
the average PSF over all 11 epochs of the observations, extracted from isolated stars in the field with brightnesses similar to that of the neighbor.
Small open circles, color-coded for each epoch, show the positions of the source relative to the bright neighbor in the detector frame at the 11 observation epochs. The source locations vary primarily because of different telescope orientations, and to a smaller extent due to different geometric distortions at the dither positions, and due to the differential PM between the two stars. 
Epochs E5 and E10 were taken at orientations differing by about $180^\circ$ from the others, but even the small change in orientation between E1 and E2 clearly places the source star either below or to the left of a bump in the superposed PSF wings of the bright neighbor.  The peak intensity of this bump is about a quarter that of the unamplified source itself. For astrometric position measurements, we use the flux distribution in a $5\times5$ WFC3 pixel box centered on the source.
The yellow square in Figure~\ref{fig:dynamic_psf} shows the size of this $5\times5$ native-pixel PSF fitting box. 

\begin{figure*}
\begin{center}
\epsscale{1.15}
\plotone{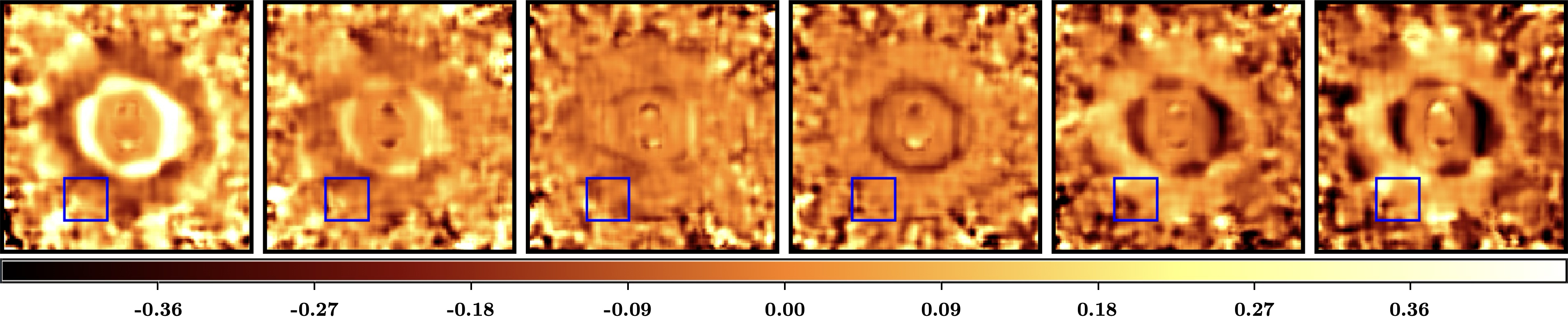}
\end{center}
\caption{
 Variations of the PSF structure in the six individual F814W images taken 
 during a single \HST\/ orbit at the E9 epoch. 
 The individual upper panels show the 
 percentage changes of the PSF with respect to the
 average PSF at each location, with a scale bar at the bottom.
 The PSF variations in the regions inside the source locations range from
 about $-35\%$  to +35\%. 
 The blue box shows the $5\times5$ native-pixel PSF fitting box centered on the location of the source star relative to its bright neighbor. 
 The gradient across the box is what affects the astrometry. 
 Using a PSF that is specifically tailored to each individual exposure removes systematic errors of $\sim 0.01$ pixels in the astrometric measurements.  
\label{fig:show_lf}
}
\end{figure*}

 As explained in S22, we take particular care to make sure that the PSF is well characterized in the wings, since the source lies in the wings of the bright star and subtracting the wings correctly is crucial for accurate astrometry.
In addition, the PSF has a strong radial gradient (along with azimuthal structure) at the location of the source.  
It is thus crucial to account for changes in the PSF caused by telescope breathing {\it within\/} each spacecraft orbit.
This is illustrated in Figure~\ref{fig:show_lf}, which shows the spatial
variations of the PSF in the six individual F814W images taken at E9.  The blue square
 is a $5\times5$ native-pixel PSF fitting box which marks the location of the source star relative to the bright neighbor. 
The percentage changes of the PSF with respect to the
 average PSF at each position within the source location range from about
 $-35\%$  to +35\%, as indicated by the scale bar at the bottom.  
 The contribution from the bright star at the source position changes from image to image, requiring the PSF subtraction to be done {\it separately for each image.}
 Using a PSF tailored specifically to each exposure removes a systematic error of $\sim$ 1\% of a pixel (corresponding to 0.4 mas) in the astrometric measurements.

\subsection{Proper Motions of Field Stars}

\begin{figure*}
\begin{center}
\plotone{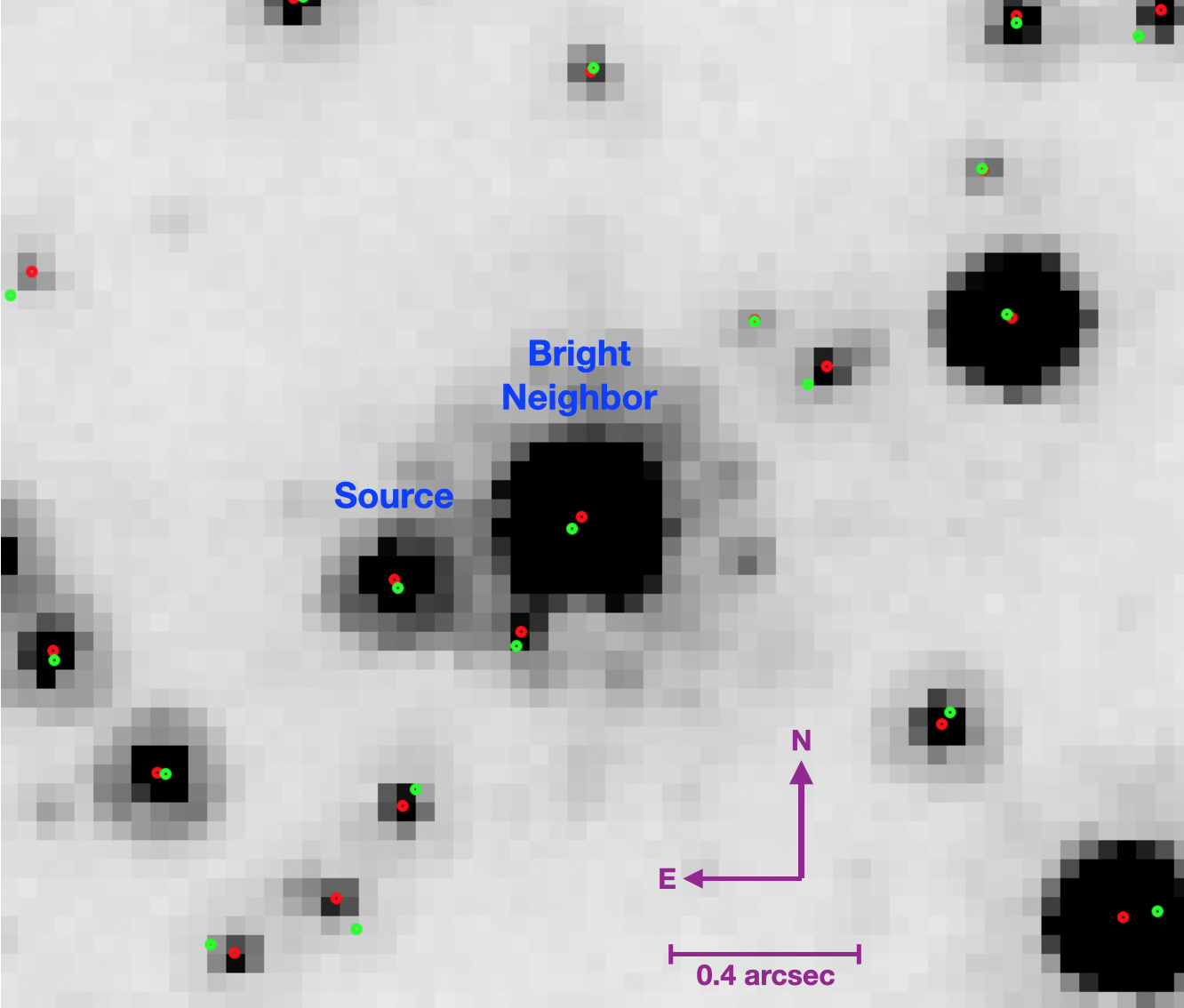}
\end{center}
\caption{
A $2\farcs4\times2\farcs0$ WFC3 F814W field showing the \OB\ field at the final epoch in 2022. The source star and its brighter neighbor are labeled.  A faint star lies just to the southeast of the neighbor, but has little effect on astrometry of the source. For each star, an open green circle shows its location at E1 in 2011, and an open red circle marks its location at E11 in 2022. Stars in this Galactic-bulge field typically move about 1 WFC3/UVIS pixel ($0\farcs040$) over the course of 11 years.
\label{fig:e11}
}
\end{figure*}

 Additionally, relative positions of the individual stars change significantly during the 11-year interval, due to their PMs. Figure~\ref{fig:e11} shows a $60\times50$-pixel ($2\farcs4\times2\farcs0$) stack of the \HST\/ images in the F814W filter at the final epoch, E11.
The source star (at nearly baseline luminosity) and its bright neighbor are labeled.  For each star in the frame, its astrometric position at the first epoch (2011) is marked by an open green circle, and its position at the final epoch (2022) by an open red circle.  In order to make accurate astrometric measurements of the source, and correctly subtract the PSF of the neighbor, we need to take these changing relative positions into account. 

\subsection{Methodology \label{subsec:methodology} }

To determine stellar positions in individual frames, we followed the procedures  described by S22.  
In brief, we used the pipeline-calibrated, CTE-corrected, un-resampled images  (filename suffixes \texttt{\_flc)} and
subtracted the PSF of the bright neighbor (see Fig.~\ref{fig:e11}) in each image using a PSF specially constructed for that particular image. We then
measured the positions and magnitudes of the source and the reference stars in the neighbor-subtracted images  (filename suffixes \texttt{\_fls)}.
We used a combination of first- and second-pass reduction techniques, based on the software tools \texttt{hst1pass}\footnote{{\tt hst1pass} is available at \url{https://www.stsci.edu/hst/instrumentation/wfc3/software-tools}.}  and
\texttt{KS2}, developed and maintained by coauthors J.~Anderson \citep{Anderson22a} and A.~Bellini \citep{Bellini17}.

We applied an iterative procedure to construct an astrometric reference frame. We started with reference stars selected to have photometric properties consistent with a location in the Galactic bulge, which were similar in color and brightness to the unamplified source star.  We chose stars that have positions and PMs cataloged\footnote{\url{https://vizier.cds.unistra.fr/viz-bin/VizieR-3?-source=I/355/gaiadr3}} in \textit{Gaia} Data Release~3 \citep[DR3;][]{Gaia2023}, and we verified that our selected stars have small parallaxes consistent with bulge membership. We then used the \Gaia\/ positions and motions to construct a reference frame for each epoch, centered on the bright neighbor star and with a plate scale of 40 mas\,pixel$^{-1}$.  Each frame was then transformed into the reference frame, using the distortion-corrected positions measured in that exposure and the positions of the reference stars at that epoch.  We then iterated several times, solving for the positions and motions of the reference stars using the \HST\/ measurements, and rejecting discrepant observations.  This resulted in a more precise reference frame than \Gaia\/ alone could produce, while at the same time ensuring that the reference frame is as close as possible to being in the absolute \Gaia\/ astrometric system.

Since the subtraction of the neighbor has a significant impact on the measured source positions,
 we explored whether reducing the fitting aperture size from 5$\times$5 pixels to 4$\times$4 or even 3$\times$3 pixels improved the fits to the lensing event.  We found consistent results no matter which of the three apertures we adopted, and retained the 5$\times$5 pixel box for the final measurements.

Since the new reductions presented here contain three additional epochs, increasing the time baseline from 6 to 11~years, 
there are improvements in the PM measurements of the reference stars. 
As a result, the new measurements of the positions are close---but not necessarily identical---to the previous values. 

\section{Modeling of Photometry and Astrometry \label{sec:fullmodeling} }

\begin{figure*}
\begin{center}
\includegraphics[width=\textwidth]{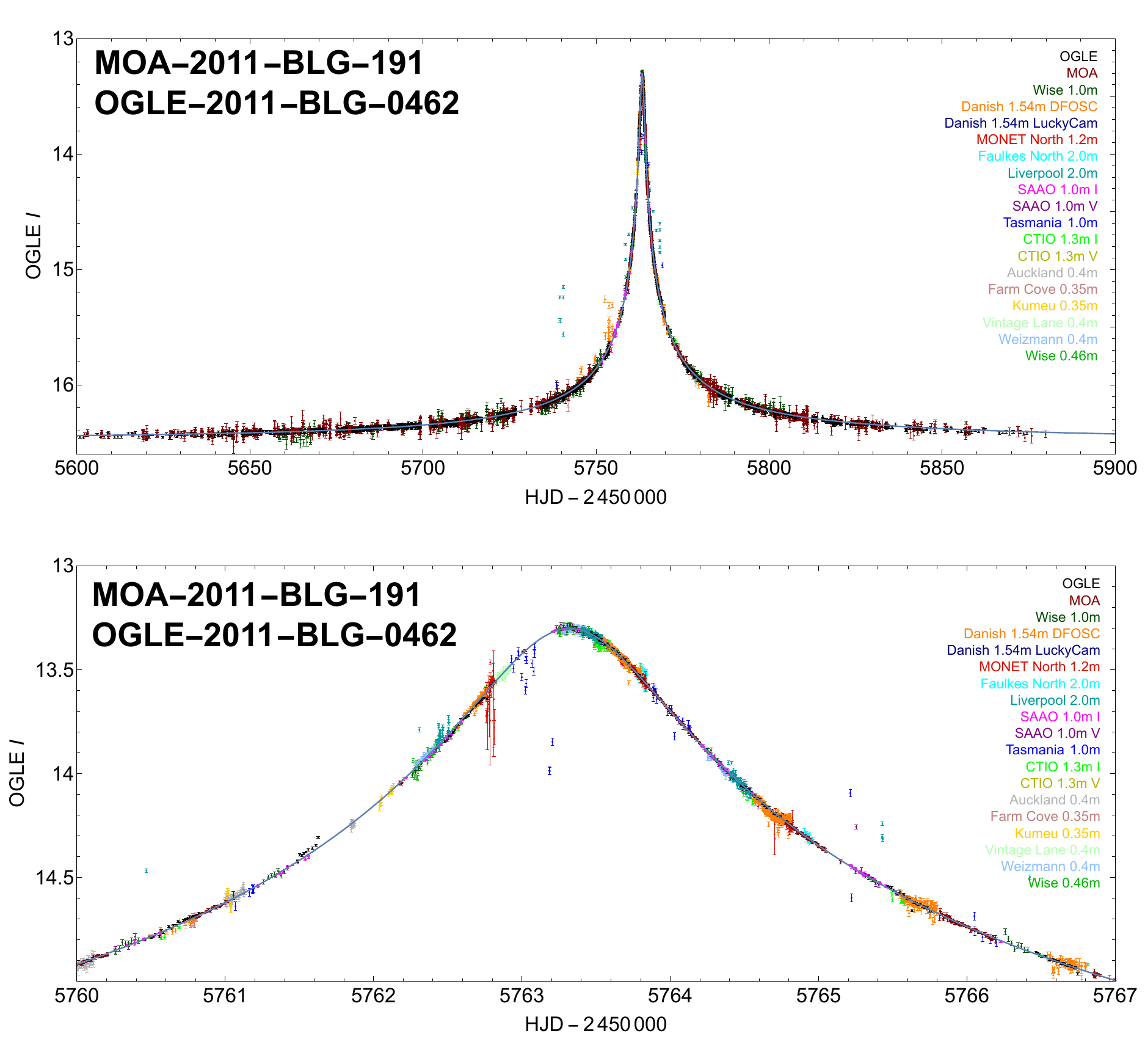}
\end{center}
\caption{Ground-based photometry of \OB\ obtained with 16 different ground-based telescopes, including the revised OGLE photometry, along with a best-fitting model lightcurve. The upper panel shows a 300-day interval, and the lower panel shows a zoomed-in version covering seven days around peak magnification. All data have been transformed to OGLE $I$ magnitudes according to the inferred baseline magnitudes and blend ratio from the common model.
\label{fig:AllPhotDataPlot}
}
\end{figure*}

The \OB\ event was monitored photometrically by 16 different ground-based telescopes, details of which are listed in Table~3 of S22. 
For our photometric analysis in the present paper we used all of the available data, 
along with the revised OGLE photometry described in Section~\ref{subsec:revised_ogle}. Given its 
largest time coverage and the highest photometric accuracy among all the 
ground-based datasets, the model fit to the microlensing lightcurve is primarily 
driven by the OGLE photometric data.
However, OGLE photometry lacks frequent sampling near the peak of the lightcurve which is 
required to provide constraints on possible binary nature of the lens. The combined photometry 
from the 16 different telescopes, many of which have frequent sampling near the peak, provide
strong constraints on possible binarity of the lens. 
Figure \ref{fig:AllPhotDataPlot} plots the measured lightcurve points, both over a 300-day interval (top panel), and zooming in on the seven days around peak magnification (bottom panel), along with a best-fitting model lightcurve. Our Figure~\ref{fig:AllPhotDataPlot} updates the similar Figure~14 in S22 by including the revised OGLE photometry. 

As noted above, there is a correlation between  \phil, the position angle of the direction of motion of the lens with respect to the source, and \piE, the microlensing parallax parameter. 
The value of \phil\ derived from photometry alone has a flat probability distribution in the range 338\textdegree $\pm15$\textdegree, with \piE\ ranging from $\sim$0.08 to 0.12.  

We carried out a combined analysis of the astrometric and photometric data, in order to obtain all of the parameters of the event. This is especially important, since the crucial parameters
$\thetaE$  (angular Einstein radius)  and \piE\ (microlensing parallax parameter) are derived from two different types of data. A simultaneous solution also provides an accurate estimate of the uncertainties in the model parameters. 

We followed the same plane-of-the-sky approach described in S22 and briefly outlined below.
We adopted a parameterization procedure wherein the model parameters contain all the terms needed to characterize the positions of the lens and the source on the sky as a function of time; these include the positions and PMs of the lens and the source, their relative parallaxes, and the angular Einstein radius of the lens.  In principle, the source parallax is also needed. However, since our reference stars and the source all belong to the bulge, the parallax of the source in our reference frame is negligible. The actual constraints on the source distance come instead from photometry and high-resolution spectroscopy, and they are described in the next section.

As described by S22, the microlensing amplification at epoch 8 was already negligible and within the photometric uncertainties of \HST\ observations (See Fig. 13 of S22). There is no further change in \HST\ photometry at epochs 9 to 11. The fit to the
photometric lightcurve provides  baseline source magnitudes of 
$m_{\rm F606W} = 21.946 \pm 0.012$ and $m_{\rm F814W} = 19.581 \pm0.012$, with corresponding blending factors for the \HST\/ photometry of 
$g= -0.012 \pm 0.015$ and $-0.006 \pm 0.012$, respectively
(where $g$ is the ratio of the flux from the neighbors included in the photometry to the flux from the unmagnified source itself).
In our subsequent calculations, we have assumed the minimum physically 
allowed value of $g=0$ for our \HST\/ photometry.
 
 The parameters of our final model are given in Table~\ref{table:parameters}, which also provides
the earlier parameters of S22 for comparison. 
We first list the fit parameters along with their uncertainties, followed by the ``derived parameters" which are calculated using the fit parameters. The value of \phil\ derived here is consistent with  \phil\ derived above from photometry alone.
The additional astrometry and revised photometry have led to reduced uncertainties and slight changes in the parameter values, which are mostly in agreement, within the uncertainties, with our previously reported values.

 Our model fit to the lightcurve is plotted as a solid line in Figure~\ref{fig:AllPhotDataPlot}. This model is used to derive physical properties of the lens, described below in Section~\ref{sec:lens_properties}.

\begin{deluxetable*}{lcccc}
  \tablecaption {Parameters from Combined Photometric and Astrometric Fits 
  \label{table:parameters} }
  \tablehead{
    \colhead{Parameter} &
    \colhead{Units} &
    \colhead{S22 Value} &
    \colhead{This Paper} &
    \colhead{Notes\tablenotemark{a}}
  }
\startdata
$ \mu_{\mathrm S} $ (RA)       & mas/yr   & 	-2.263  $\pm$   0.029 &    -2.210   $\pm$   0.023	 & (1)  \\
$ \mu_{\mathrm S} $ (Dec)      & mas/yr   & 	-3.597  $\pm$  0.030 &    -3.533    $\pm$   0.024	 & (1)  \\
$ \theta_{\mathrm E} $         & mas      & 	 5.18	$\pm$   0.51  &	5.496     $\pm$   0.310  & (2) \\
$ t_{\mathrm E}^\star $              & days     &    270.7	$\pm$  11.2   &   273.1 $\pm$  6.3   & (3)  \\
\phil                      & deg      &    342.5	$\pm$   4.9   &   336.7 $\pm$   3.9  & (4)  \\
$ t_0^\star $ (HJD$-$2450000.0)& days     &   5765.00	$\pm$   0.87  &  5765.99  $\pm$   0.78   & (5)  \\
$ \pi_{\mathrm E} $      &          & 	 0.0894 $\pm$   0.0135&	0.0944        $\pm$   0.0124	 & (6)  \\
$ u_0^\star $           &          & 	 0.0422 $\pm$   0.0072&	0.0492        $\pm$   0.0065	 & (7)  \\   
Blending parameter             &          & 	18.80	$\pm$   0.79  &    18.89  $\pm$   0.65 & \\
Derived parameters:&&&&\\
$u_0$                  &          &  0.00271		  &  0.00271	        	      & (8) \\
$ t_0$  (HJD-2450000.0)        & days     &  5763.32		  &  5763.32	        	      & (9) \\
\enddata 
\tablenotetext{a}{Notes: (1) Undeflected PM of the source in \Gaia\/ DR3 absolute frame; 
  (2) Angular Einstein radius;
  (3) Angular Einstein radius $\thetaE$ divided by absolute value of lens-source PM $\mu_\mathrm{LS}$;
  (4) Orientation angle of the lens PM relative to the source (N through E);
  (5) Time of closest angular approach without parallax motion;
  (6) Relative parallax of lens and source in units of $ \theta_\mathrm{E} $; 
  (7) Impact parameter in units of $ \theta_\mathrm{E} $ without parallax motion;
  (8) Impact parameter derived using the model-fit parameters above and after including parallax motion, in units of $\theta_\mathrm{E}$;
  (9) Time of closest angular approach derived as above and after including parallax motion.
  }
\end{deluxetable*}

\begin{figure*}
\begin{center}
\plotone{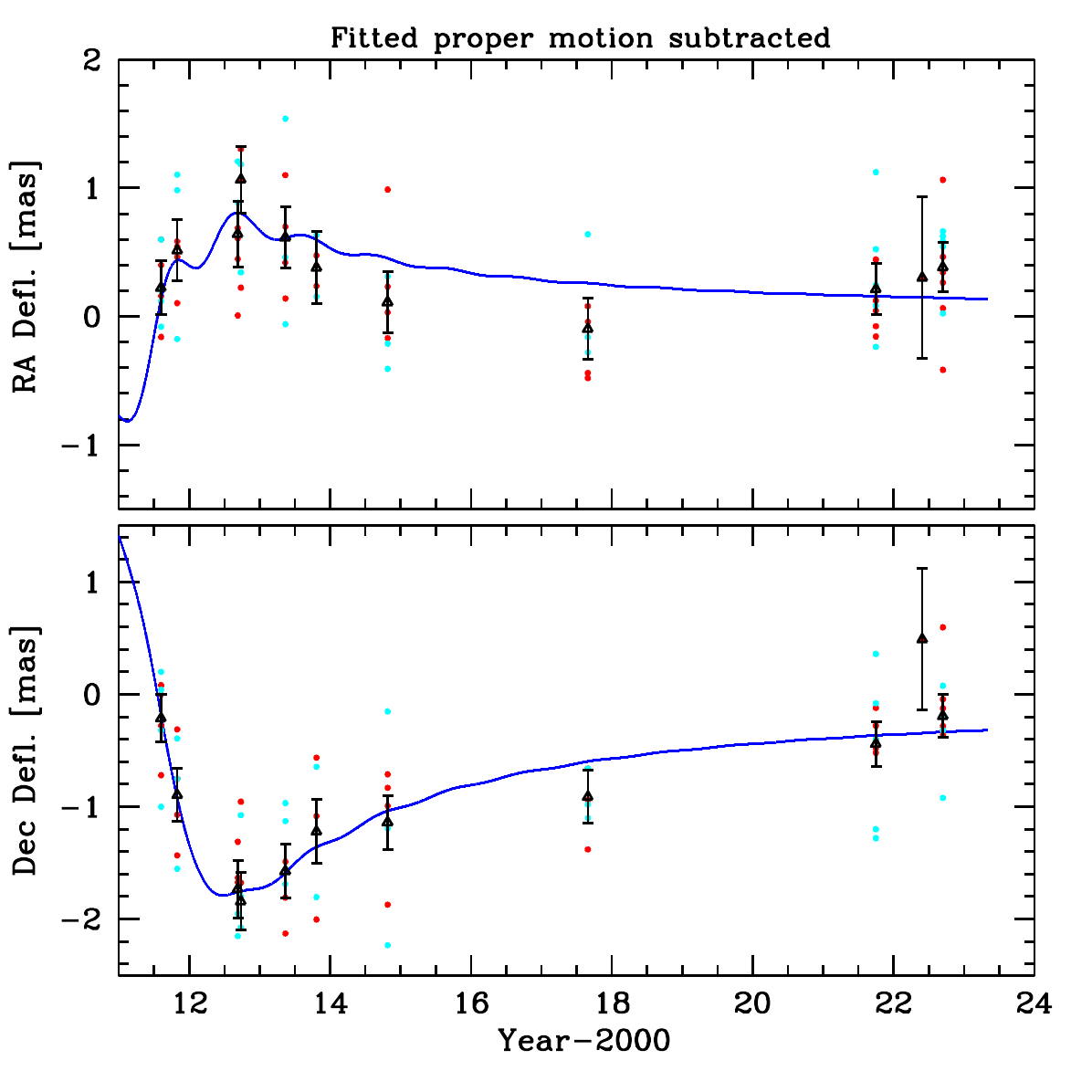}
\end{center}
\caption{Predicted and measured positions of the source for our adopted joint astrometric and photometric fit. The fitted PM for the source has been subtracted from both model and measurements, in order to allow a better scaling of the plot. Small dots show the individual measurements from {\HST\/} images (cyan for F606W, red for F814W). There is no systematic difference between the average measurements in 
F814W and F606W filters. The black triangles with error bars show the averages of all the
measurements at each \HST\/ epoch, with the associated uncertainties.  The blue line represents our adopted model.  
\label{fig:xy_deflections_vs_time}
}
\end{figure*}

In Figure~\ref{fig:xy_deflections_vs_time} we plot the measured and predicted source positions separately for the R.A.  and decl. coordinates. 
The fitted linear PM of the source has been subtracted from both model and data, and therefore the plotted points and curves directly represent the relativistic deflection of the apparent source position.  The blue line is our final adopted model fit to the data. It's worth noting here that, on statistical grounds, $\sim$32\% of the data points are 
expected to fall outside of $1\sigma$ from the model fit, which is consistent with what is observed. 
Additionally, there is no $>2\sigma$ deviation for any of the 22 data points. 
Thus the deviations and the distribution of the errors are consistent with
theoretical expectations. 

\begin{figure*}
\begin{center}
\plotone{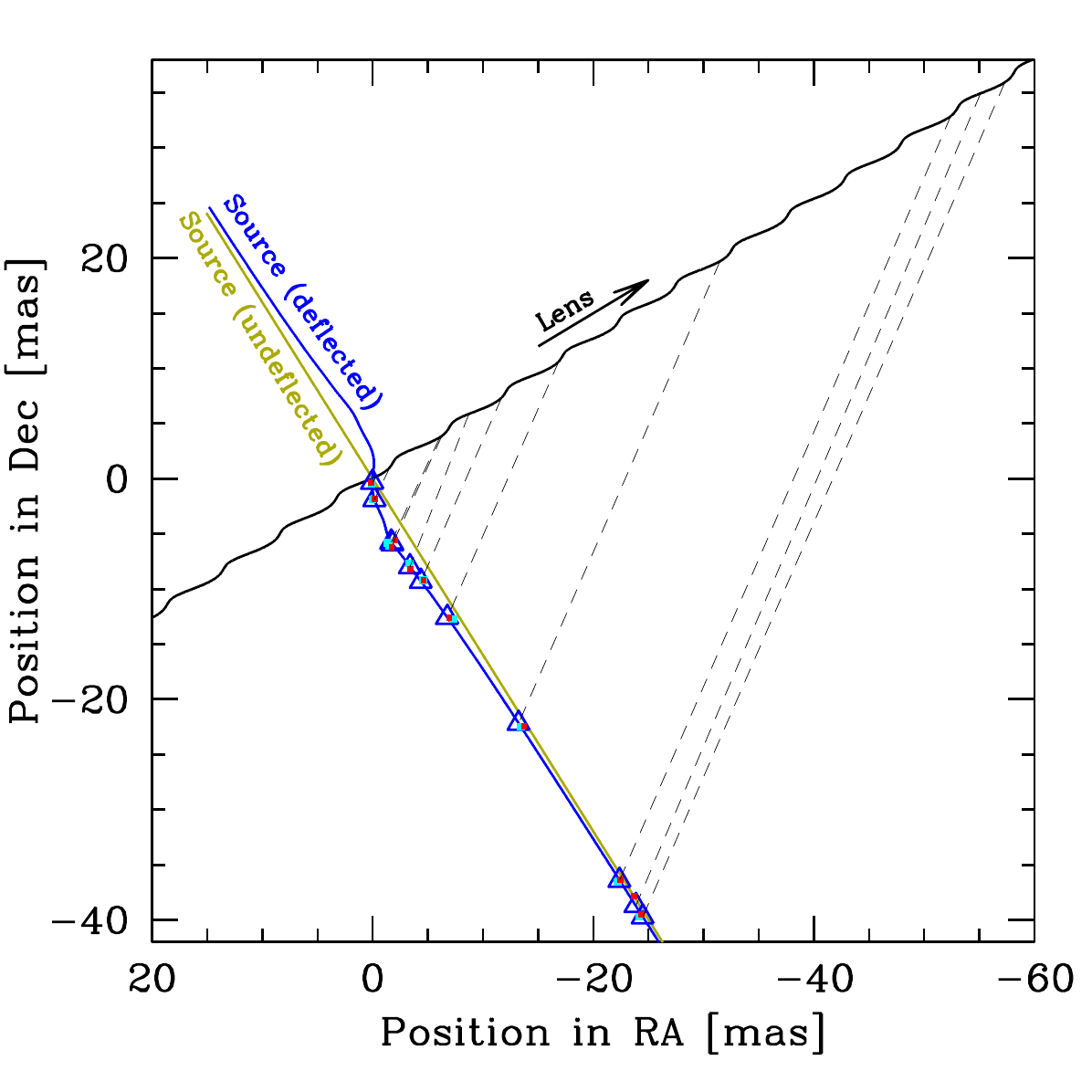}
\end{center}
\caption{Reconstructed motions of the lens and of the source on the plane of the sky.  The reference point is the undeflected position of the source at the time of closest approach $ t_0 $.  The source is moving from the upper left to lower right, and the lens is moving from lower left to upper right as shown. The small cyan and red squares are the average measurements from {\HST\/} images in F606W and F814W filters at each epoch.  For clarity, the individual 
measurements are not shown here, but they are shown in Fig.~\ref{fig:xy_deflections_vs_time}, along with the error bars. The blue triangles are the predicted positions at each epoch from the model.  The dashed gray lines connect the undeflected source and lens positions at each {\HST\/} epoch.}
\label{fig:plane_of_sky}
\end{figure*}

Figure~\ref{fig:plane_of_sky} maps the reconstructed motions of the lens (black line) and the undeflected source (dark green line) in the plane of the sky, based on our final model. The ``wobble'' of the lens's path is due to parallax.
The predicted deflected apparent source trajectory is shown by the blue solid line. The mean astrometric measurements at each \HST\/ epoch in F606W and F814W are shown as small filled cyan and red squares, respectively. Blue open triangles show the model position at each epoch. Dashed gray lines connect the model lens position to the undeflected and deflected source positions at each epoch.

\section{Updated Source Distance \label{sec:source_distance} }

As described below, the mass determination for the lens does not depend upon the individual distances to the lens and to the source, but only on the relative lens-source parallax, $\piLS$, and the Einstein ring radius, $\thetaE$ (see S22 for further details). These quantities are directly determined from the microlensing lightcurve and the measured astrometric deflections. However, we still need to know the distance to the source, $\DS$, in order to determine the distance to the lens, $\DL$; this is of astrophysical interest, as discussed in the next section.  Here we update the  distance to the source.

In the S22 study, we adopted the photospheric parameters of the source star determined by \citet{Bensby2013} from high-resolution spectra obtained near the light-curve maximum ($\Teff=5382\pm92$~K, $\log g=3.80\pm0.13$, and $\rm[Fe/H]=+0.26\pm0.14$), and placed it in the distance-independent $\log g$ versus $\log\Teff$ plane (Figure~19 in S22). By comparison with two independent sets of theoretical stellar isochrones (``PARSEC'' and ``BaSTI''; see S22 for details and references) we obtained the corresponding
absolute magnitudes (Vegamag scale) for the source star in the native \HST/WFC3 bandpasses. The results from the two isochrone sets agreed very well with each other, giving means 
of $M_{\rm F606W}=+3.025$ and $M_{\rm F814W}=+2.390$, and thus an intrinsic color of $(m_{\rm F606W}-m_{\rm F814W})_0 = 0.635$. 
Using the observed baseline color of $m_{\rm F606W}-m_{\rm F814W} = 21.946 - 19.581 = 2.365$ (S22, Table~1), we found a color excess of $E(m_{\rm F606W}-m_{\rm F814W})=1.73$.

In our earlier analysis, we had converted this color excess to the absorption at F814W, $A_{\rm F814W}$, by adopting the interstellar-extinction formalism of \citet{Cardelli1989} and \citet{ODonnell1994}, and using a ``standard'' value of $R_V=A_V/E(B-V)=3.1$ for the ratio of visual total to selective absorption. However, several studies indicate that a 
lower value of $R_V=2.5\pm0.2$ is more appropriate
for interstellar extinction in the  direction toward the Galactic bulge \citep[e.g.,][]{Nataf2013}. 
Moreover, we had used the values of $A_\lambda/A_V$ given for the \HST/WFC3 filters at the PARSEC website\footnote{\url{http://stev.oapd.inaf.it/cmd}}; however, these are calculated for a star with the spectral-energy distribution (SED) of a G2\,V star, whereas our target is a considerably redder source. 

To improve the extinction calculation, we obtained a model-atmosphere SED from the \citet{Castelli2003} compilation\footnote{Conveniently available at \url{http://svo2.cab.inta-csic.es/theory/newov2}.} for a star with $\rm[Fe/H]=+0.2$ lying at the grid point nearest to the source parameters, at $(\Teff, \log g)=(5500\,{\rm K}, 4.0)$. We also downloaded\footnote{From \url{https://www.stsci.edu/hst/instrumentation/wfc3/performance/throughputs}.} the throughput curves for the WFC3 UVIS2 detector and the F606W and F814W filters. We then convolved these  functions with various amounts of interstellar extinction, using the Cardelli-O'Donnell  algorithms with $R_V=2.5\pm0.2$, to find the amount of extinction that reproduces the observed $E(m_{\rm F606W}-m_{\rm F814W})$ color excess. The corresponding reddening is $E(B-V)=2.13\pm0.10$, giving an extinction of $A_{\rm F814W} = 2.93\pm0.18$~mag. This is significantly smaller than the 3.33~mag found by S22, due to the smaller value of $R_V$ and the longer effective wavelengths for the cool and reddened source. Both of these changes reduce the calculated amount of extinction, thus increasing the inferred distance of the star.

Correcting the measured F814W baseline magnitude for the source by applying this newly derived extinction, and using the absolute F814W magnitude given above, we find a distance of $D_S=7.1\pm1.7$~kpc. The quoted error is the formal uncertainty (dominated by the uncertainty in the spectroscopic $\log g$), but there are likely additional systematic errors, given the large amount of extinction and uncertainties in the interstellar-extinction function toward the Galactic bulge (see the review by \citealt{Nataf2016}). We conclude that the source star is a late-type subgiant, lying in the Galactic bulge, possibly on the side nearer to us.\footnote{
In the calculations reported by M22, those authors adopted the source distance from S22 of $D_S=5.9\pm1.3$~kpc. However, in their penultimate paragraph, they mention a 
larger distance of $8.8\pm1.4$~kpc. This was based on an absolute magnitude in the ground-based $I$ bandpass of $M_I=+2.83$, inferred by \citet{Bensby2013} from theoretical isochrones, a baseline $I$ magnitude, and the assumption that the extinction of the source is the same as that of the red clump of Galactic bulge stars at the source site. Ground-based photometry of this source at baseline suffers from a large blending correction due to the neighboring brighter star, as discussed in S22 and the present paper. We prefer to use the cleaner flux measurements from \HST\/ data and to determine the reddening directly for the source itself.}

\section{Properties of the Lens \label{sec:lens_properties} }

In this section, we discuss the physical properties of the lensing object of the \OB\ microlensing event, based on our updated analysis.  These properties
are summarized in Table~\ref{table:lensproperties}.

\begin{deluxetable*}{lcc}
\tablecaption{Parameters of the \OB\ Black Hole Lens \label{table:lensproperties} }
\tablehead{
\colhead{Property} &
\colhead{Value} &
\colhead{Sources \& Notes\tablenotemark{a}} 
}
\startdata
J2000 right ascension, $\alpha$      & 17:51:40.208  & (1) \\
J2000 declination, $\delta$          & $-29$:53:26.50 & (1) \\
Galactic coordinates, $(l,b)$        & $359\fdg86, -1\fdg62$ & (1) \\
Mass, $M_{\rm lens}$                      & $7.15 \pm 0.83\, M_\odot$               & (2) \\ 
Distance, $D_L$                           & $1.52 \pm 0.15$ kpc                   & (3) \\
PM rel to source, $(\mu_\alpha, \mu_\delta)_{\rm rel}$ & $(-2.907\pm 0.145, +6.751\pm 0.336) \rm\,mas\,yr^{-1}$ & (4) \\
Absolute PM, $(\mu_\alpha, \mu_\delta)_{\rm abs}$ & $(-5.117\pm 0.254, +3.217\pm 0.161) \rm\,mas\,yr^{-1}$ & (5) \\
PM rel to local stars, $(\mu_\alpha, \mu_\delta)_{\rm local}$ & $(-5.47\pm 0.53, +4.51\pm 0.43) \rm\,mas\,yr^{-1}$ &  (6) \\
PM rel to local stars, $(\mu_l, \mu_b)_{\rm local}$ & $(+1.05\pm 0.13, +7.01\pm 0.66) \rm\,mas\,yr^{-1}$ & (7) \\
Transverse velocity, $(V_l, V_b)_{\rm trans,local}$ & $(+7.59 \pm 0.97, +50.49 \pm 4.73) \kms$ & (8)\\
Total transverse velocity, $V_{\rm trans,local}$& $51.1 \pm 7.5 \kms$ & (9) \\
\enddata
\tablenotetext{a}{Sources \& notes: (1) S22, Table~1; (2) From \S\ref{subsec:mass}; (3)~From \S\ref{subsec:lensdistance}; (4) PM of lens relative to source in equatorial coordinates; (5)~Absolute PM of lens in \Gaia\/ J2000 frame; 
(6)~PM of lens relative to mean of \Gaia\/ stars at similar distances in equatorial coordinates; (7)~PM of lens relative to nearby stars in Galactic coordinates; (8)~Transverse velocity components relative to local stars in Galactic coordinates; (9)~Total transverse velocity relative to local stars.}
\end{deluxetable*}

\subsection{Lens Mass \label{subsec:mass} } 

The mass of the \OB\ lens is determined from the values of its angular Einstein radius, $\thetaE$, and the microlensing parallax, $\pi_{\rm E}$ (see S22). 
Using the values from Table \ref{table:parameters},
we obtain a mass of  
\begin{equation}
M_{\rm lens} \equiv  \frac{\theta_\mathrm{E}}{\kappa \pi_\mathrm{E}} = 7.15 \pm 0.83\,M_\odot \, , 
\label{eq:mass}
\end{equation}
\noindent where $\kappa \equiv 4 G / [c^2 ({\rm1 \, AU})] \simeq 8.144 \, {\rm mas} \ M_\odot^{-1}$.

\smallskip

\subsection{Lens Distance \label{subsec:lensdistance} }

The distance to the lens is determined from
\begin{equation}
\piLS \equiv \pi_\mathrm{E}\, \thetaE \equiv 
\pi_\mathrm{L}-\pi_\mathrm{S} \equiv (1~\mbox{AU})  \left(\frac{1}{D_\mathrm{L}} - \frac{1}{D_\mathrm{S}}\right) \, 
\end{equation}
Using the values of 
\piE\ and $\thetaE$ as above, and the distance to the source of $\DS=7.1\pm1.7$~kpc from Section~\ref{sec:source_distance}, we derive a lens distance of
\begin{equation}
\DL= 1.52\pm0.15\rm\,kpc \, .
\end{equation}

\smallskip

\subsection{Lens Motion \label{subsec:lens_motion}}

The PM components of the lens relative to the source in equatorial coordinates, derived from the parameters in 
Table \ref{table:parameters}, are given in the sixth row of Table~\ref{table:lensproperties}. Using the absolute PM of the source in the \Gaia\/ frame, from the first two rows in Table \ref{table:parameters}, we obtain the absolute PM components of the lens, given in the seventh row of Table~\ref{table:lensproperties}.

This absolute PM can be converted to an absolute transverse velocity, using the known distance to the lens. However, a more physically meaningful quantity is the velocity of the lens with respect to the 
stars in its surrounding neighborhood. 
The median \Gaia\/ PMs of Galactic-disk stars at distances similar to that of the lens, as derived by S22, are
$0.35 \pm 0.61$ and $-1.29 \pm 0.52\,\rm mas\, yr^{-1}$ in RA and Dec, respectively. 
The PM components of the lens relative to the local stars are given in row eight. In row nine these are converted to PMs in the directions of Galactic longitude and latitude.  Row ten then uses the lens distance to convert these to the relative transverse velocities in Galactic coordinates.

The final row in
Table \ref{table:lensproperties} gives the total transverse space velocity of the BH relative to the mean of the local stars (which have similar Gaia parallactic distance as the BH), which is 
\begin{equation}
V_{\rm trans,local} = 51.1 \pm 7.5 \kms \, .
\end{equation}
This is larger than the velocities of almost all local stars (see Figure~\ref{fig:show_log} below). It supports the suggestion that the BH received a modest natal kick during its supernova explosion, under the assumption that the progenitor arose from this local population.

\subsection{Limits on Lens Luminosity \label{subsec:lens_luminosity} }

As shown in row six of Table \ref{table:lensproperties}, the PM of the lens with respect to the source is $\sim$$7.35\rm\,mas\,yr^{-1}$. Thus the lens would be separated from the source by $\sim$47 and $\sim$84 mas at the times of the \HST\/ observations obtained in 2017 and 2022, respectively. S22 did not find any excess light at the expected location of the lens at the 2017 epochs, placing a limit on the lens apparent magnitude of $I>24.6$. (Here we make the approximation of equating the \HST\/ 
$m_{\rm F814W}$ magnitude to the $I$ magnitude in the ground-based system.) 
The larger expected separation between the lens and the source at the final 2022 \HST\/ epoch allows us to place 
deeper limits on the lens's optical luminosity. 
In order to search for any excess light from the lens, we selected PSFs from an online library\footnote{\url{https://www.stsci.edu/hst/instrumentation/wfc3/data-analysis/psf}}; these were subtracted from the E11 F814W images (in the six individual un-resampled {\tt \_flc} frames).  We saw no indication of any light at the lens position, placing a limit on the lens apparent magnitude of $I>25.1$.
This corresponds to the luminosity of a main-sequence star of mass $\sim$0.15\,\msun. 
Thus our derived mass of $7.15 \pm 0.83$\,\msun, coupled with the lack of detected light at the lens location, confirms that the \OB\ lens must be a BH\null. 

There have been some attempts \citep[e.g.,][]{2022Chmyreva} to use the brightness limit of $I>24.6$ 
to place constraints on the accretion luminosity from the BH\null. Our revised limit to the $I$-band luminosity of $I>25.1$ thus provides a stronger limit to the accretion luminosity
\citep{Kimura2025}.

\subsection{Limits on Stellar Companions of the Lens}

As noted in Section~\ref{subsec:firstblackhole}, all of the stellar-mass BHs discovered up to the present time, except for \OB\ itself, have been in binary systems. Most of these binaries have been so close together that they are interacting X-ray sources, or even binaries that merged and emitted gravitational radiation. More recently, however, \Gaia\/ astrometric data have revealed three BHs orbiting normal stars in non-interacting binaries with wider separations, with periods of 0.5, 3.5, and 11.6~years \citep{2023MNRAS.518.1057E, 2023AJ....166....6C, 2023MNRAS.521.4323E, Gaia2024}.  For a review of these findings, see \cite{2024NewAR..9801694E}. 

\begin{figure}
\begin{center}
\plotone{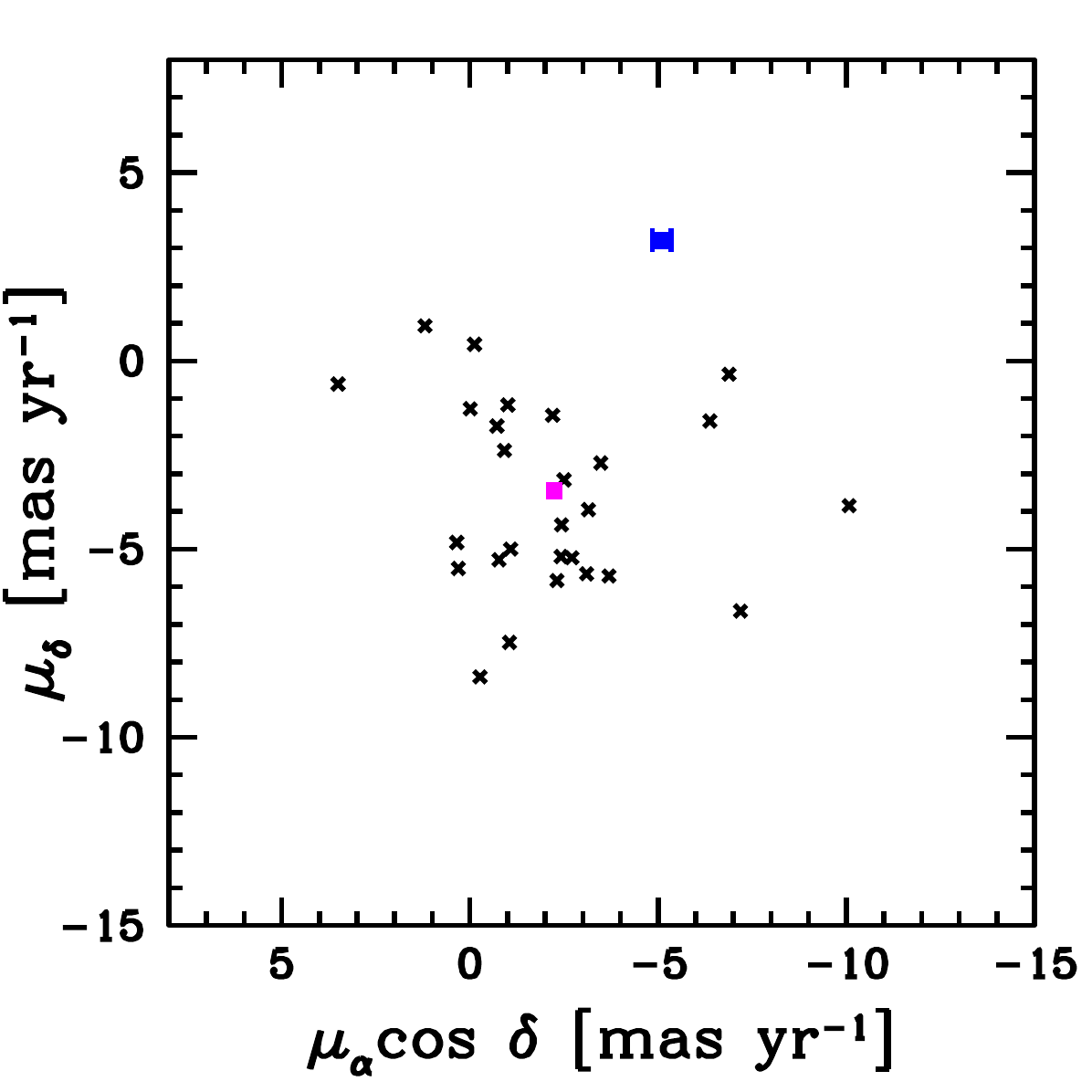}
\end{center}
\caption{Our measured PMs of all the stars brighter than $I\sim\,24.5$ within an angular radius of $1\farcs 3$ from the BH\null. The errors are of the order of 1 mas yr$^{-1}$ at the faint end of $I\sim\,24.5$ and smaller for the brighter stars. The source and the lens are shown in magenta and blue, respectively. None of the stars within this field have PMs similar to the BH. 
\label{fig:show_log}
}
\end{figure}

A similar close companion to \OB\ lying within about 230~AU is already ruled out by the absence of deviations from a point-source--point-lens lightcurve in the dense photometric sampling of the \OB\ event (see S22). 
Moreover, as discussed in Section~\ref{subsec:lens_luminosity}, there is no detectable light at the location of the BH at late times, and there is no star visible within 0\farcs 2 of the lens. This rules out any unresolved main-sequence companions down to $0.15\,M_\odot$, at a separation of up to $\sim$300~AU. 

However, a significant fraction of massive stars are known to have stellar companions at 
large orbital separations, extending to several thousand AU \citep[e.g.,][]{2019igoshev, Hwang2020}.
It is therefore of interest to investigate whether \OB\ could be accompanied by a distant, {\it resolved\/} stellar companion. 
Such a gravitationally bound companion would have a similar PM to that of the BH\null. 

We note that a companion at $>$2000 AU from the BH would have an orbital motion of $< 2\kms$. With an expected kick velocity as discussed in the previous Section, 2000 AU should be a safe upper limit beyond which a stellar companion is unlikely to survive. To search for a possible wide companion, we examined the PMs of all stars detected in our \HST\/ frames within an angular radius of $1\farcs 3$, corresponding to 2000~AU at the distance of the lens, centered on the BH.

Figure \ref{fig:show_log} shows  the PMs of all the stars 
down to $I\simeq24.5$ lying within  $1\farcs 3$ of the BH (black points). 
Note that the PM measurement requires high S/N, so the limiting magnitude is
brighter than the limit for simple detection.
The source and lens are plotted with magenta and blue points, respectively. The PM of the BH is well separated from PMs of all the stars in the surrounding $1\farcs 3$ region (black points), showing that none of the stars in the field share a common PM with the BH\null. 
This rules out any companion down to $\sim$0.2\,\msun\  within 2000 AU of the BH.  

\begin{deluxetable*}{llcl}
\tablewidth{0pt}
\tablecaption{Limits on Binary Companions of the Lens \label{table:binarity} }
\tablehead{
\colhead{Technique}           &
\colhead{Result} &
\colhead{Orbital}&
\colhead{Companion Mass Limit}\\
\colhead{}           &
\colhead{} &
\colhead{Separation}&
\colhead{}
}
\startdata
Microlensing lightcurve & No anomaly & $<$230 AU &No companion $>$0.2\msun  \\
Lens luminosity & No detection at $I<25.1$& $<$300 AU &No stellar companion  $>$0.15\msun  \\
Proper motions  & No co-moving star at $I<24.5$ & $<$2000 AU &No stellar companion $>$0.2\msun  \\
\enddata
\end{deluxetable*}

Thus, our search for a binary companion did not lead to any detection. The results of our search are summarized in Table \ref{table:binarity}, where we provide the distances and mass limits of a possible binary companion, obtained through different methods.

\section{Discussion and Critique of Previous Work}

\begin{figure*}
\begin{center}
\epsscale{1.17}
\plotone{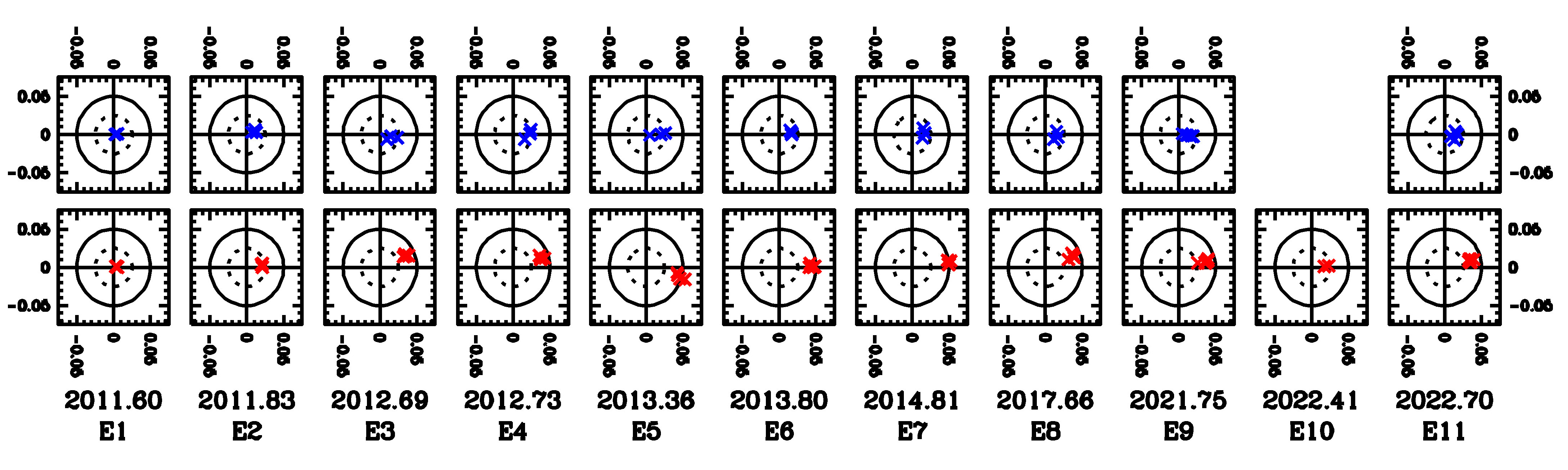}
\end{center}
\caption{
We measured the locations of the source in the original pipeline-processed {\tt flc} images and in modified {\tt fls} images, which had the bright neighbor subtracted.  The crosses show the difference between these two measurements for individual exposures, shown for all the 11 epochs (left to right). The top row shows the F606W observations and the bottom row shows F814W.  The dotted circle corresponds to a correction of 1 mas, and the solid circle to 2 mas. 
The F814W corrections are naturally greater due to more extended PSF at redder wavelengths, and correspondingly larger contamination from the bright neighbor.
}
\label{fig:pos_corr}
\end{figure*}

In Table \ref{table:useddata} we summarize the data used and the properties explored in different studies. The major differences are: (i) for photometry, we use photometric data collected from 16 different observatories, while all other studies use only the OGLE data, (ii) for astrometry, we carry out separate bright-star PSF subtractions for every image before measuring the source positions while L23 use a single ``bias correction"  (correction applied to account for the effect of the bright neighbor) for each epoch from source injection/recovery, 
(iii) we use spectroscopic measurements to derive the source distance, while other studies use only
the photometry, and (iv) we investigate in detail the possible binarity of the BH.

\begin{deluxetable*}{lcccccc}
  \tablecaption {Data Used and Properties Explored in Different Studies   \label{table:useddata}}
  \tablehead{
    \colhead{Parameter} &
    \colhead{S22 } &
    \colhead{L22 DW/EW$^*$} &
    \colhead{M22 } &  
    \colhead{L23} &
    \colhead{This paper} & 
    \colhead{Notes$^a$}
    }
\startdata
 Ground-based & Data from 16	& Old OGLE  & Updated OGLE &  Updated OGLE & Data from 16  \\
Photometry& telescopes including &data only &data only &data only& telescopes including  \\
&  old OGLE data&&&&updated OGLE data\\
\\
 \HST\/ data 	& 8 epochs& 8 epochs & Used L22/S22& 11 epochs& 11 epochs\\
  &&& measurements \\
\\
Bright neighbor& PSF subtraction & Source injection/ & Used L22/S22 & Source injection/ & PSF subtraction\\
correction & (in individual images& recovery & measurements& recovery& (in individual images \\
&to account for & (single value && (single value & to account for \\
&\HST\/ breathing) & per epoch)&&per epoch)& \HST\/ breathing)\\
\\
Astrometric & Measurements of  & Only averages& & Only averages& Measurements of\\
Measurements & all individual images&for each epoch&&for each epoch& of all individual images\\
 \\
 Spectroscopy     &  VLT spectroscopy &None & None& None& VLT spectroscopy& (1)\\
 \\
 PMs of local stars    & From Gaia& None& From galaxy model & None & From Gaia & (2)\\
 \\
 Source distance& Using CMD & -- & Adopted & Red& Using CMD \\
 &and spectroscopy &&S22&clump?& and spectroscopy& (3)\\
 \\
 Possible Binarity  & Using dense  	&None & None & None & Dense photometry \\
 & photometry at peak&&&& at peak and PMs\\
 &&&&&of nearby stars\\
 \\
 Lens luminosity & Using PSF subtraction & No PSF & --& No PSF& PSF subtraction\\
 & at lens location& subtraction&&subtraction\\
\enddata
\tablenotetext{a}{Notes: (1) VLT spectra are used for distance determination of the source (2) The PMs of local stars are used to estimate the velocity of the
BH with respect to the local stars. (3) L23 quote a distance of 8.8 $\pm$ 1.4 kpc for the source. They do not cite its origin, but this is the same as the distance mentioned in the penultimate paragraph of M22 as described in Section \ref{sec:source_distance}. 
}
\end{deluxetable*}

In Table \ref{table:comparison} we summarize the physical parameters of the \OB\ event derived by the investigations published to date, along with our new analysis.
The L22 values had the largest discrepancies which are now superseded by L23, 
but they are included here for completeness. Most of the parameters in L23 are in broad agreement
with those found in the present paper (and in M22). 
Nevertheless, some discrepancies are evident: for example, the value of $\thetaE$ and hence the lens mass, remains consistently smaller 
in L23 and have larger error bars. The velocities of the lens are also discrepant (smaller) compared to the present study.
This is worth exploring further, since L23 used not only the same datasets for astrometry, but also the same astrometry software developed by our team
\citep{Bellini17, Anderson22a}, as described in Section~\ref{subsec:methodology}. Here we discuss possible causes of the remaining discordances. 

\begin{deluxetable*}{lcccccc}
  \tablecaption {Physical Parameters from Different Studies$^{*}$   \label{table:comparison}}
  \tablehead{
    \colhead{Parameter} &
    \colhead{Units} &
    \colhead{S22 } &
   \colhead{L22 DW/EW$^*$} &
    \colhead{M22 } &  
    \colhead{L23} &
    \colhead{This paper} 
  }
\startdata
$ \mu_{\mathrm S} $ (RA)   & mas/yr   &-2.263   $\pm$ 0.029    & -2.25/-2.25     &-2.281   $\pm$   0.022& -2.02   $\pm$   0.01					  & -2.210   $\pm$ 0.023	\\
$ \mu_{\mathrm S} $ (Dec)  & mas/yr   &-3.597   $\pm$ 0.030    &	-3.57/-3.56  & -3.624   $\pm$  0.021& -3.45   $\pm$   0.02					  & -3.533   $\pm$ 0.024	\\
$ \theta_{\mathrm E} $     & mas      & 5.18 $\pm$ 0.51  &  3.89/4.13      &5.68 $\pm$   0.40 & $4.79^{+1.13}_{-1.15}$ &  5.496 $\pm$ 0.310  \\
 \phil                  & deg      &342.5 $\pm$4.9  &   355.47/18.08  &  342.3 $\pm$ 3.0 	&  $343.75^{+4.80}_{-3.95}$    & 336.7 $\pm$3.9  \\
$ \pi_{\mathrm E} $  &	      & 0.0894 $\pm$ 0.0135	   &  0.12/0.24      &0.095    $\pm$   0.009	&0.10 $\pm$ 0.01						  &  0.0944 $\pm$ 0.0124	\\
M$_L$                      &M$_\odot$ &7.1 $\pm$ 1.3  	   &   3.79/2.15     &  7.88	 $\pm$ 0.82	&  $6.03^{+1.19}_{1.04}$	  & 7.15 $\pm$ 0.83		\\
$\DL$\                       & kpc      & $1.45 \pm 0.15$ 	   &   1.67/0.92     &  $1.49 \pm 0.12$ 	&  $1.72^{+0.32}_{-0.23}$	  &  $1.52 \pm 0.15$		\\
$ \mu_{(\mathrm L, abs)} $ (RA)   & mas/yr   & -4.36 $\pm$ 0.22   & -2.64/-0.69    &-4.48  $\pm$   0.39& 
$-3.80^{+0.48}_{-0.55}$			  & $-5.117\pm 0.254$ \\
$ \mu_{(\mathrm L, abs)} $ (Dec)  & mas/yr   &3.06 $\pm$ 0.66 &	1.46/1.53 & 3.29   $\pm$  0.50& 2.60$^{+0.83}_{-0.80}$				  & 3.217$\pm$ 0.161\\
V$_{T,L}$                   &$\kms$    &45 $\pm$ 5 	   & 23.95/7.26 & 43.4 $\pm$ 3.8		& $\sim 40.0 \pm 7.6 ^{\dagger}$&51.1 $\pm$ 7.5 	\\
$I_L^\mathrm{\ddagger}$ & mag & $>24.6$ & -- & -- & -- & $>25.1$ \\
\enddata
\tablenotetext{*}{L22 values are superseded by L23, but they are included here for completeness.
DW refers to their ``Default Weight" solution and EW refers to ``Equal Weight"}
\tablenotetext{\dagger}{L23 only provide the absolute velocity. In order to express the motion with respect to local stars, we have used our measurement of the average velocity of local stars described in \ref{subsec:lensdistance}}
\tablenotetext{\ddagger}{Limit on the apparent $I$-band magnitude of the lens, derived from 
non-detection of light at the expected lens position.}
\end{deluxetable*}

First, we used photometric data from 16 different ground-based telescopes, whereas L23 used only the OGLE photometric data. The main cause for the discrepancy comes from 
different approaches to accounting for how the unrelated bright star about 10 pixels from the source affects the astrometric and photometric measurement of the source. L22 and L23 measured this effect through injection of stars similar to the source star close to a few other neighboring stars, using the {\tt KS2} software routine in L22 and {\tt hst1pass} in L23.  
 As described in section Section~\ref{subsec:methodology}, here (and in S22), we used a custom procedure for extracting an extended PSF and then subtracting that PSF from each exposure before measuring the source star.  This model PSF went out to a radius of 15 pixels, far enough to completely cover the region where the source star was measured.  The PSF was extracted from stars similar in color and brightness to the bright, interfering neighbor.  A separate model PSF was constructed for each individual exposure, since the telescope breathing can impact the location and intensity of the bumps in the PSF halo.  It's worth pointing out here that in their Fig.~12, L23 specify a single value of a ``bias correction" for all epochs in the S22 work. This is misleading since, as explained in detail above and in S22, we treated each image separately and did not use any ``bias correction.”

Thus, the procedure we used to account for the presence of the bright neighbor was quite different from that in L22 and L23.  Figure \ref{fig:pos_corr} shows the residual between the uncorrected and corrected measurement for the source.  The units are 40 mas WFC3/UVIS pixels.  The inner circle is drawn at 0.025 pixel (1 mas) and the outer circle at 0.05 pixel (2 mas).  It is clear that the correction is small when the source is bright (early epochs) and larger when it is fainter.

Our corrections and measured deflections are clearly larger than those of L22 and L23, particularly in the R.A. direction, based on inspection of Figures 4 and 9 of L23. But it is hard to make a direct comparison, because we derive a separate correction and an associated error value for each image at each epoch, while those studies provide only a single value for each epoch. We emphasize that constructing a correction based on inserted stars near other neighbors (as done in L22 and L23) requires a precise knowledge of the true position of the source star in each epoch relative to its neighbor, which then necessitates an iterative procedure. From Figure 1, it is clear that the impact of the bright star depends very specifically on where the source star lands relative to the bright star’s bumpy PSF extended features. Insertion over a range of locations cannot account for this, while our extended PSF modeling and subtraction naturally accounts for the relative locations of the source with respect to the PSF features and removes the neighbor contribution from the exact location. 

\vskip 1cm

\section{Summary}

We have reanalyzed the dataset of \OB\ taking into account all the photometric data collected from 16 different telescopes,  including the revised OGLE photometry. We have used the three additional epochs of \HST\/ observations taken in 2021 and 2022 which increases the baseline of \HST\/ observations from 6 to 11 years. This leads to better accuracy in the PM measurements of all the stars in the field including the source, resulting in more accurate measurements of the deflections. In the earlier analysis, there was a slight tension between the direction of motion of the lens derived from photometry and astrometry. The tension disappears after taking the revised OGLE photometry with updated errors into account. Our analysis suggests that the BH has a mass of $7.15 \pm 0.83\, M_\odot$, lies at a distance of $1.52\pm0.15$~kpc, and is moving with a space velocity of $51.1 \pm 7.5 \kms$ relative to stars in its neighborhood. These results are in overall agreement with the other studies, all of which now agree that \OB\ is undoubtedly a BH\null. We explore the reasons for the remaining  discrepancy and show that the small scale PSF variation and \HST\/ breathing are the main causes, which we have taken into account in order to correct for the contribution of the nearby bright star. We show that the BH has no main-sequence stellar companion with a mass of more than $\sim$$0.2\,M_\odot$ at any orbital separation. 

{\it \bf Acknowledgements:} Based in part on observations with the NASA/ESA {\it Hubble Space Telescope\/} obtained at the Space Telescope Science Institute, which is operated by the Association of Universities for Research in Astronomy, Incorporated, under NASA contract NAS5-26555. 
We thank Przemek Mr\'oz and Andrzej Udalski for updating the OGLE data, and Casey Lam and Jessica Lu for obtaining the additional HST data used in this paper under proposal ID 16760.
The HST data presented in this article were obtained from the Mikulski Archive
for Space Telescopes (MAST) at the Space Telescope Science Institute. The
specific observations analyzed can be accessed via \dataset[doi:10.17909/v97n-kh49]{https://doi.org/10.17909/v97n-kh49}.



\bibliography{ms.bib}{}

\bibliographystyle{aasjournal}

\end{document}